\documentclass[aps,prl,twocolumn,superscriptaddress,amsmath,amssymb,showpacs]{revtex4-1}
\usepackage{amssymb}
\usepackage{graphicx}
\usepackage{bm}
\usepackage{subfigure}
\usepackage{epsfig}
\usepackage{color}
\usepackage[ansinew]{inputenc}

\def\3he{{$^3${\rm He}}}

\def\slD{\raise.15ex\hbox{$/$}\kern-.57em\hbox{$D$}}
\def\dsl{\raise.15ex\hbox{$/$}\kern-.57em\hbox{$\Delta$}}
\def\slp{{\raise.15ex\hbox{$/$}\kern-.57em\hbox{$\partial$}}}
\def\nsl{\raise.15ex\hbox{$/$}\kern-.57em\hbox{$\nabla$}}
\def\sla{\raise.15ex\hbox{$/$}\kern-.57em\hbox{$\rightarrow$}}
\def\slla{\raise.15ex\hbox{$/$}\kern-.57em\hbox{$\lambda$}}
\def\gtwid{\raise.3ex\hbox{$>$\kern-.75em\lower1ex\hbox{$\sim$}}}
\def\ltwid{\raise.3ex\hbox{$<$\kern-.75em\lower1ex\hbox{$\sim$}}}

\def\12{{1\over2}}

\def\part{\partial}

\def\bk{{\bf k}}
\def\br{{\bf r}}
\def\bR{{\bf R}}

\def\bethlogo{\vbox{\bf \line{\hrulefill} 
    \kern-.5\baselineskip 
    \line{\hrulefill\phantom{ ELIZABETH A. MASON }\hrulefill} 
    \kern-.5\baselineskip 
    \line{\hrulefill\hbox{ ELIZABETH A. MASON }\hrulefill} 
    \kern-.5\baselineskip 
    \line{\hrulefill\phantom{ 1411 Chino Street }\hrulefill} 
    \kern-.5\baselineskip 
    \line{\hrulefill\hbox{ 1411 Chino Street }\hrulefill} 
    \kern-.5\baselineskip 
    \line{\hrulefill\phantom{ Santa Barbara, CA 93101 }\hrulefill} 
    \kern-.5\baselineskip 
    \line{\hrulefill\hbox{ Santa Barbara, CA 93101 }\hrulefill}
    \kern-.5\baselineskip 
    \line{\hrulefill\phantom{ (805) 962-2739 }\hrulefill} 
    \kern-.5\baselineskip 
    \line{\hrulefill\hbox{ (805) 962-2739 }\hrulefill}}}
\def\lisalogo{\vbox{\bf \line{\hrulefill} 
    \kern-.5\baselineskip 
    \line{\hrulefill\phantom{ LISA R. GOODFRIEND }\hrulefill} 
    \kern-.5\baselineskip 
    \line{\hrulefill\hbox{ LISA R. GOODFRIEND }\hrulefill} 
    \kern-.5\baselineskip 
    \line{\hrulefill\phantom{ 6646 Pasado }\hrulefill} 
    \kern-.5\baselineskip 
    \line{\hrulefill\hbox{ 6646 Pasado }\hrulefill} 
    \kern-.5\baselineskip 
    \line{\hrulefill\phantom{ Santa Barbara, CA 93108 }\hrulefill} 
    \kern-.5\baselineskip 
    \line{\hrulefill\hbox{ Santa Barbara, CA 93108 }\hrulefill}
    \kern-.5\baselineskip 
    \line{\hrulefill\phantom{ (805) 962-2739 }\hrulefill} 
    \kern-.5\baselineskip 
    \line{\hrulefill\hbox{ (805) 962-2739 }\hrulefill}}}

\def\low#1{\lower.5ex\hbox{${}_#1$}}
\def\ltwid{\raise.3ex\hbox{$<$\kern-.75em\lower1ex\hbox{$\sim$}}}

\def\psl{\raise.15ex\hbox{$/$}\kern-.57em\hbox{$\partial$}}
\def\partt{\raise.15ex\hbox{$\widetilde$}{\kern-.37em\hbox{$\partial$}}}
\def\parts{\raise.15ex\hbox{$/$}{\kern-.6em\hbox{$\partial$}}}
\def\nablas{\raise.15ex\hbox{$/$}{\kern-.6em\hbox{$\nabla$}}}
\def\oprod{\hbox{$\rm O$}{\kern -0.8em\hbox{$\Pi$}}}
\def\partw#1{\raise.15ex\hbox{$/$}{\kern-.6em\hbox{${#1}$}}}

\def\si{{\sigma}}

\def\gtappr{{{\lower4pt\hbox{$>$} } \atop \widetilde{ \ \ \ }}}
\def\ltappr{{{\lower4pt\hbox{$<$} } \atop \widetilde{ \ \ \ }}}

\def\topppageno1{\global\footline={\hfil}\global\headline
={\ifnum\pageno<\firstpageno{\hfil}\else{\hss\twelverm --\ \folio
\ --\hss}\fi}}

\def\toppageno2{\global\footline={\hfil}\global\headline
={\ifnum\pageno<\firstpageno{\hfil}\else{\rightline{\hfill\hfill
\twelverm \ \folio
\ \hss}}\fi}}

\def\ltdash{\raise-1.8pt\hbox{$\scriptscriptstyle |$}}

\newlength{\upit}\upit=0.1truein

\newlength{\bxwidth}\bxwidth=1.5 truein

\newcommand{\dg}{^{\dagger }}

\newcommand{\pmat}[1]{\begin{pmatrix} #1 \end{pmatrix}}

\newlength{\figwidth}
\newlength{\shift}
\newlength{\fight}

\newcommand \bea {\begin{eqnarray} }
\newcommand \eea {\end{eqnarray}}

\newsavebox{\fmbox}

\newcommand{\figgy}[3]{
\begin{figure}[t]
\centering
\mbox{\includegraphics[width=\figwidth]{#1}}
\caption{#2}\label{#3}
\end{figure}
}
\newcommand{\ybal}{$\beta$-YbAlB$_4\,$}
\newcommand{\aybal}{$\alpha$-YbAlB$_4$\ }

\begin{document}

\title{$\beta$-YbAlB$_{4}$: a critical nodal metal}

\author{Aline Ramires}
\affiliation{Department of Physics and Astronomy, Rutgers University, Piscataway, New Jersey, 08854, USA}

\author{Piers Coleman}
\affiliation{Department of Physics and Astronomy, Rutgers University, Piscataway, New Jersey, 08854, USA}
\affiliation{Department of Physics, Royal Holloway, University
of London, Egham, Surrey TW20 0EX, UK.
}
\author{Andriy H. Nevidomskyy}
\affiliation{Department Physics and Astronomy,
Rice University, Houston, Texas, 77005, USA}

\author{A. M. Tsvelik}
\affiliation{Department of Condensed Matter Physics and Materials Science,
Brookhaven National Laboratory, Upton, New York, 11973, USA}

\date{\today}

\begin{abstract}
We propose a model for the intrinsic quantum criticality of
$\beta$-YbAlB$_4$, in which a vortex in momentum space gives rise to a new
type of Fermi surface singularity.  The unquenched angular
momentum of the $|J=7/2, m_J=\pm5/2\rangle$ Yb 4$f$-states
generates a momentum-space line defect in the hybridization 
between 4$f$ and conduction electrons, leading
to a quasi-two dimensional Fermi surface with a
$k_{\perp }^{4}$ dispersion and a singular density of states
proportional to $E^{-1/2}$.  We discuss the implications
of this line-node in momentum space for our current understanding of
quantum criticality and its interplay with topology.
\end{abstract}

\maketitle

Since their discovery, heavy fermion materials have provided
a wealth of insights into correlated electron physics. These
materials contain a matrix
of localized magnetic moments formed from $f$-electrons
immersed within a host metal; at low temperatures the
spin-quenching 
entanglement of the $f$-moments with the conduction electrons
gives rise to a diversity of low energy ground-states, including
anisotropic superconductors, Kondo insulators and
quasiparticles with effective masses 100s of times 
that of bare electrons \cite{coleman,heavyfermion2}. An important sub-class of heavy fermion materials
exhibit the phenomenon of quantum criticality, whereby the metal can be tuned via pressure, doping or magnetic field through a
zero temperature second-order quantum phase transition, 
around which they show
non-Fermi Liquid behavior and predisposition to superconductivity
\cite{sci,Gen,Ste}.

The  discovery \cite{Nak,Mat} of an intrinsically quantum
critical heavy fermion superconductor \ybal has attracted much
interest. \ybal exhibits
quantum criticality without tuning: in the absence of a field it 
exhibits non-Fermi liquid behavior, with a $T^{3/2}$
temperature dependence of the resistivity and a $T^{-1/2}$ divergence
in the magnetic susceptibility, but 
a magnetic field induces an immediate cross-over into a Fermi liquid  (FL)
with a $T^{2}$ resistivity, in which the susceptibility diverges as $B^{-1/2}$. Intriguingly,  $T/B$ scaling of the
free energy and magnetization has been observed over 4 decades in
magnetic field \cite{Mat}, pin-pointing the critical magnetic field to
within 0.1 mT of zero while demonstrating 
that the Fermi temperature of the field-induced FL is 
the Zeeman energy\cite{Mat}.

In this paper, we show that the  properties of this material can be
understood in terms of a \textit{nodal} hybridization model. In essence
\ybal resembles a Kondo insulator, but one in which the 
the hybridization gap between the conduction
and f-electrons vanishes along a line in momentum space,
producing a critical semi-metal with a singular density of
states. 

There are three known
examples of such nodal materials: CeNiSn, CeRhSb and CeCu$_{4}$Sn, in
which the hybridization appears to vanish linearly along a line in
momentum space, closing the gap to form a heavy fermion semi-metal
\cite{IM}.  
In \ybal the unusual local seven-fold symmetry of the
ytterbium (Yb) site surrounded by boron (B) atoms protects a
``high-spin'' $\vert J\!=\!7/2,\, m_J\!=\!\pm 5/2\rangle $ state, in
which the $f$-electrons carry a large unquenched orbital angular
momentum.  The $\vert5/2\rangle$ state carries at least two units of
unquenched orbital momentum orientated along the $c$-axis, yet plane
waves carry no orbital angular momentum in the direction of motion, so
the $f$-state is protected from hybridization with conduction
electrons traveling along the $c$-axis.  This causes the
hybridization to develop a singular structure, $V (\bk )\sim (k_{x}\pm
i k_{y})^{2}$ vanishing as the square of the transverse momentum
$\bk_{\perp}$ with a double-vorticity associated with the two
unquenched units of orbital angular momentum.  The electrons and holes
at the band-edge then form an emergent two-dimensional electron gas
with a dispersion proportional to the square of the hybridization,
\begin{equation}\label{}
E (\bk )\sim
|V(\bk)|^2 \sim (k_{\perp })^{4},
\end{equation}
giving rise to a quartic dispersion with
a divergent density of states $N (E)\propto E^{-1/2}$.
It is the
field-induced doping of this two-dimensional heavy band that accounts
for the unusual field-tuned behavior in \ybal.

\figwidth=5cm
\figgy{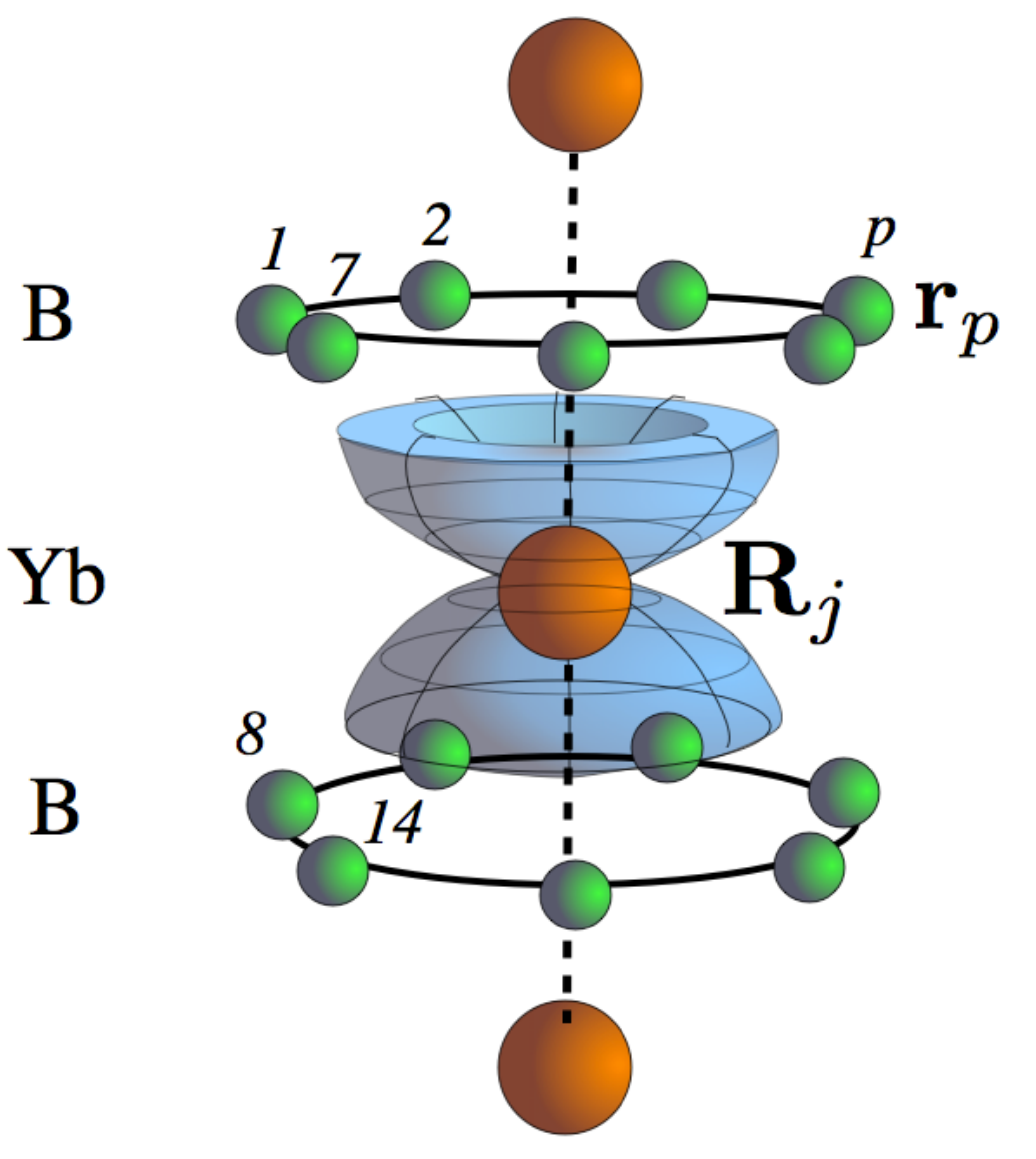}{Showing the seven-fold symmetric
environment of the Yb$^{3+}$ ions (large spheres) in \ybal, sandwiched between two
heptagonal rings of $B$ atoms (small spheres). The blue surface is the orbital distribution in the $m_j=\pm5/2$ state.}{fig1}

In $\beta$-YbAlB$_4$ the Yb atoms form a honeycomb lattice, sandwiched
between layers of B atoms, with the Yb atoms sitting between a pair of
7-member B rings, giving rise to a local environment with local
seven-fold symmetry \cite{Nak}, as shown in Fig.~\ref{fig1}.  
We shall assume that the Yb ions are in a nominal Yb$^{3+}$, $4f^{13}$
configuration, with total angular momentum $J=7/2$. 
Photoemission spectroscopy indicates a 
microscopic valence of $2.75$ \cite{ybal-valence} due to 
moment-conserving valence fluctuations 
$Yb^{3+}\leftrightarrow Yb^{2+}+e^{-}$. 

$J\!\!\!=\!\!\!7/2$ crystal field operators with  7-fold and time-reversal
symmetries conserve  total $J_z$, splitting the J=7/2 
Yb multiplet into four
Kramers doublets, each with definite $|m_J|$.
The  Curie constant and the Ising anisotropy of the magnetic
susceptibility  of \ybal  are  consistent with
a pure Yb ground-state doublet \hbox{$|J=7/2,m_J=\pm5/2\rangle$} \cite{Nev}, a configuration
that exhibits maximal
hybridization with the seven-fold boron rings. 
This Ising ground-state is also consistent with the large anisotropic g-factor
observed in electron spin
resonance measurements on \ybal \cite{Pag,Ram}.

We model the low energy
physics of \ybal as a layered Anderson lattice \cite{Nev},
\begin{eqnarray}
&&H=\sum_{n,k,\sigma} \epsilon_{\bk n}c\dg_{\bk n\sigma }c_{\bk n\sigma}
+\sum_{j}H_{m} (j),
\end{eqnarray}
where the first term describes a tight-binding
boron conduction electron band where $n$ is the band index, and
\begin{equation}\label{}
H_m (j) = V_{0} (c\dg_{j\alpha }X_{0\alpha } (j)+ {\rm
h.c.})+E_{f}X_{\alpha \alpha } (j),
\end{equation}
describes the hybridization with the Yb ion at site $j$ and the energy level $E_f$ of the $f$-electrons. Here,
$X_{0\alpha } = \vert 4f^{14}\rangle \langle 4f^{13},\alpha
\vert $ is a Hubbard operator linking the
$4f^{13}$, $m_{J}\equiv \alpha = \pm 5/2$ state of the Yb$^{3+}$ ion to the
completely filled shell Yb$^{2+}$ state $\vert 4f^{14}\rangle $.
The operator
\begin{eqnarray}\label{calpha}
c\dg _{j\alpha }=
\sum_{p\in(1,14),\sigma}
c\dg _\sigma(\textbf{R}_{jp})
\
\mathcal{Y}_{\sigma\alpha }(\textbf{r}_{p}),
\end{eqnarray}
creates a conduction electron in a Wannier state delocalized across the
seven-fold boron rings directly above and below the Yb ion
at site $j$,
with local $f$
symmetry and $J_{z}= \alpha = \pm 5/2$.
The  $\bf R_{jp}= \bf R_{j}+ {\bf r}_{p}$ are the locations of the
fourteen boron sites around the Yb site $j$ (see Fig. 1).
The hybridization matrix,
\begin{eqnarray}
\mathcal{Y}_{\sigma\alpha }(\textbf{r})=C_{\sigma \alpha}^{\frac{7}{2}}
Y^{3}_{\alpha -\sigma }(\textbf{r})
= \frac{1}{\sqrt{7}}\pmat{
\sqrt{6}Y^{3}_{2}
&
Y^{3}_{3} \cr Y^{3}_{-3} & \sqrt{6}Y^{3}_{-2}
}(\hat \br ),
\end{eqnarray}
where the $C_{\si\alpha}^{\frac{7}{2}}=\langle 3 \alpha -\sigma,
\frac{\sigma}{2} \vert \frac{7}{2},\alpha \rangle $ are
Clebsch-Gordan coefficients for the Yb$^{3+}$,  $\alpha = \pm 5/2$ configurations.

We employ
a slave boson decomposition of the Hubbard operators, $X_{0\alpha
} (j) = b_{j}\dg f_{j\alpha }$, where $b_{j}$ and $f_{j\alpha }$ are
a slave boson and an Abrikosov pseudo-fermion respectively;
in a mean-field approximation, 
\begin{eqnarray}
H_{m}(j)\!=\!  V_{0}^{*}[c\dg_{j\alpha }f_{j\alpha } + {\rm h.c.}]+
\tilde{E}_{f}f_{j\alpha}^\dagger f_{j\alpha}
+
\lambda_{0} (r^{2}-1),
\end{eqnarray}
where $V_{0}^{*}$ is the quasiparticle hybridization, renormalized by
the mean-field amplitude of the
slave boson field,
$r=|\langle b\rangle| $
taken to be constant at each site. $\lambda_{0}$ imposes the mean-field constraint $\langle n_{f}\rangle
+ r^{2}=1$, while the renormalized position of the $f$-level $\tilde{E}_{f}= \lambda_{0}+E_{f}$.

Next, we transform to
momentum space and evaluate the form-factor of the seven-fold
symmetric Yb-B cluster.
To obtain a simplified model, let us assume a single band 
of dispersion $\epsilon_{\bk }$ hybridizing with the Yb atom.
Rewriting the creation operator at a given boron site
in terms of a plane wave state $c\dg_{\sigma } ({\bf R}_{jp})
={(4\cal N)}^{-1/2}\sum_{\bk }c\dg_{\bk
\sigma }
e^{-i\bk \cdot \bR_{jp}}
$,  and $f_{j\alpha }= {\cal
N}^{-1/2}
\sum_{\bk }f_{\bk\alpha }e^{i\bk \cdot \bR_{j}}
$
where ${\cal N}$ is the number of
Yb sites, Eq.~(\ref{calpha}) becomes:
\[
c\dg_{j\alpha } = {(4\cal N)}^{-1/2}
\sum_{\bk  \sigma }c\dg_{\bk\sigma}\gamma_{\sigma \alpha}(\bk)e^{-i\bk \cdot \bR_{j}},
\]
where the form-factor of the Yb-B cluster
\begin{equation}\label{formfactor}
[\underline{\gamma}(\bk )]_{\sigma\alpha  } =
\sum_{p=1,14}
\mathcal{Y}_{\sigma\alpha }(\textbf{r}_{p})e^{-i\bk \cdot {\bf r}_{p}}.
\end{equation}

The mean-field
Hamiltonian (6) can then be written in terms of the plane-wave
$c_{\bk \sigma}$ and
$f_{\bk \alpha}$ operators as
\begin{eqnarray}\label{heff}
H_{eff}=\sum_{\bk } (c\dg_{\bk },f\dg_{\bk })
\pmat{\epsilon_{\bk }\mathbb{I} & \underline{V} (\bk )\cr
\underline{V}\dg (\bk )& \tilde{E}_{f}\mathbb{I}}
\pmat{c_{\bk }\cr f_{\bk }},
\end{eqnarray}
where all details of the hybridization are hidden in the matrix
$[\underline{V}(\textbf{k})]=\frac{1
}{2}V_0^{*}
\underline{\gamma}(\bk )
$.
Now in polar co-ordinates
\begin{eqnarray}
\mathcal{Y}(\hat \br) = \sqrt{\frac{5}{64 \pi}}s_{\theta}^2\pmat{
6 c_{\theta } e^{2i \phi }& - s_{\theta } e^{3i\phi }\cr s_{\theta } e^{-3 i\phi }& 6 c_{\theta } e^{-2 i \phi
} },
\end{eqnarray}
where we denote $(\cos \theta ,\sin \theta)\equiv (c_{\theta
},s_{\theta })$.
The important point, is that 
the hybridization vanishes as $\sin^{2}\theta$ along the
$c$-axis.
Now the effect of Fourier transforming in Eq.~(\ref{formfactor}), is 
to replace the
real-space argument by the momentum
${\cal
Y} ( \br)
\rightarrow {\cal Y} ( \bk) $.
To obtain an analytic expression, we approximate the discrete sum over the  positions in the seven-fold B ring by a continuous integral:
$\sum_{p}\rightarrow 7
\sum_{\pm}\int \frac{d\phi }{2 \pi}$.
We find that  $V (\bk )$ is proportional to a unitary matrix,
\begin{equation}\label{}
V (\bk ) = i \tilde{V}_{0}\pmat{\alpha_{\bk } & \beta_{\bk } \cr
-\beta_{\bk } ^{*}& \alpha_{\bk} ^{*}},
\end{equation}
where $\tilde{V}_{0}= \frac{7 V_{0}^{*}}{16}\sqrt{\frac{5}{\pi}}$ and
\begin{eqnarray}\label{l}
\alpha_{\bk } &=& 6 \sin (k_{z}a/2)
 (\hat  k_{x}+i \hat
k_{y})^{2} J_{2} (k_\perp R)\cr
\beta_{\bk } &=& \ \ \cos(k_z a/2)
 (\hat  k_{x}+i \hat
k_{y})^{3}J_{3}(k_\perp R),
\end{eqnarray}
where $J_n$ are Bessel functions of order $n$, $R$ is the radius of the seven-fold rings and
$a$ is the distance between boron layers.
Since $J_{n} (x)\propto x^{n}$ at small $x$, 
near the $c$-axis, the hybridization vanishes as
$k_\perp^2$,
with a diagonal form
\[
V (\bk )\sim \pmat{(k_{x}+ik_{y})^{2}& 0\cr
0 & (k_{x}-i k_{y})^{2}}.
\]

As one proceeds around the $c$-axis, the phase of the hybridization
advances  by $4\pi$, forming a 
double vortex in the hybridization
along the $c$-axis. This
 \textit{vorticity} is a consequence of angular momentum conservation 
about the $c$-axis: plane waves $\vert \bk\sigma\rangle$
traveling along the $c$-axis carry a spin
angular momentum of $\pm \frac{1}{2}$ along the $c$-axis, and because
the $f$-states are in an $m_{J}=\pm \frac{5}{2}$, angular momentum
conservation prevents the mixing of conduction and $f$-electron waves
travelling along the $c$-axis.

\figwidth=7.5cm
\figgy{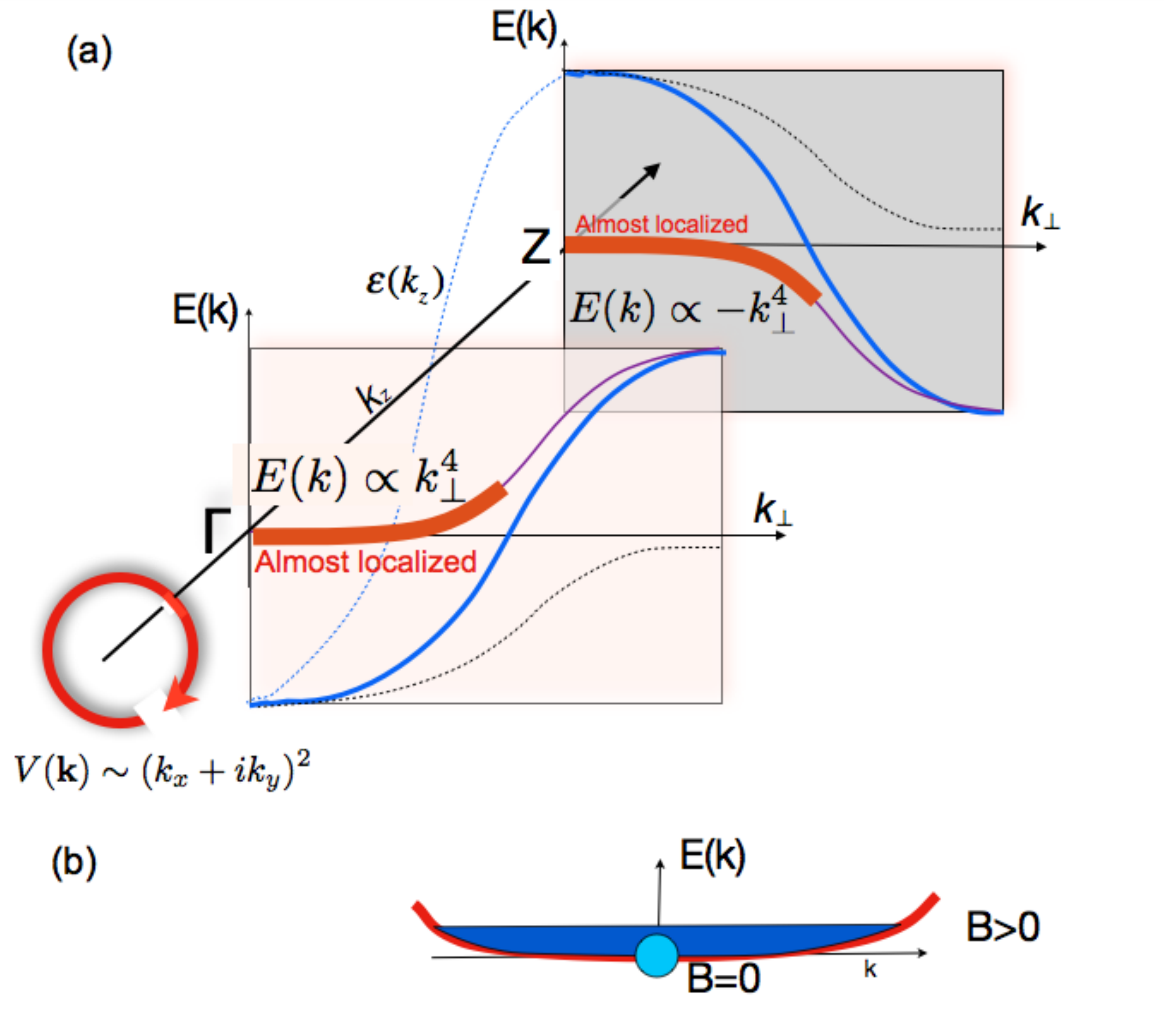}{(a) Showing dispersion around the $c$-axis,
with an electron pocket at the $\Gamma$ point and a hole pocket at the
$Z$ point. (b) Magnetic field fills the $k_{\perp}^{4}$ band.
}{hybrid}

We can diagonalize the mean-field Hamiltonian, to obtain
a hybridized dispersion
\begin{eqnarray}
E^{\pm}_{\bk }= \frac{1}{2} (\epsilon_{\bk }+\tilde{E}_{f})\pm
\bigl [\frac{1}{4} (\epsilon_{\bk }-\tilde{ E}_{f})^{2}
+|V(k)|^2
\bigr]^\frac{1}{2}
,
\end{eqnarray}
where $|V (k)|^{2}= \tilde{V}_{0}^{2} [ |\alpha_{\bk }|^{2}+ |\beta_{\bk }|^{2} ]$.
Fig. \ref{hybrid} illustrates the hybridized band-structure.  Near the $c$-axis,
the squared hybridization vanishes as $V (\bk )^{2}= A (k_{z}) k_{\perp
}^{4}$.
The dispersion in the vicinity of the $c$-axis is then given by
\[
E (k_{\perp },k_{z})=\tilde{E_{f}} + \frac{V (\bk )^{2}}{-\epsilon
(k_{z})}\approx \tilde{E}_{f} + \eta (k_z)
k_{\perp }^{4}.
\]
where $\eta (k_{z})= \frac{A (k_{z})}{-\epsilon
(k_{z})}$
and we have assumed that $|\epsilon (k_{z})|$ is large compared to $|V(k)|$.
In other words, the system develops an emergent two-dimensional Fermi surface,
with a $k_{\perp }^{4}$ dispersion. A hole band is formed in the
region where $\epsilon (k_{z})>0$, while an electron band
is formed in the region where $\epsilon (k_{z})<0$.
In the case where  $\epsilon
(k_z)$ changes sign along the $c$-axis, a two dimensional electron and
hole band is formed above and below the $f$-level.

To explain the intrinsic criticality of \ybal we conjecture that the
$f$-level is pinned to zero energy $\tilde{E}_{f}=0$.  A heuristic
argument for this assumption, is to regard \ybal as a Kondo insulator
in which the nodal hybridization closes the gap along the
$c$-axis, pinching the $f$-level in the gap at precisely zero energy.
At the current stage of understanding, this assumption is purely
phenomenological, a point we return to later.

If $\tilde{E_{f}}=0$, the density of states for this
dispersing system is then given by $N^{*} (E)= \sum_{\pm } N_{\pm
}^{*} (E)\theta (\pm E )$, where
\begin{equation}\label{}
N_{\pm}^{*} (E) =  2\int k_{\perp }\frac{dk_{\perp }}{dE_{\pm}}
\frac{dk_{z}}{(2\pi)^{2}}=
\frac{1}{\sqrt{|E| T_{0}^{\pm }}}
\end{equation}
where $\frac{1}{\sqrt{T^{\pm}_{0}}} = \frac{1}{8 \pi^{2}}
\int \frac{dk_{z}}{\sqrt{|\eta (k_{z})|}}\theta [\mp\epsilon (k_{z})]$
determines the characteristic scales $T^{\pm }_{0}$
for the electron (+) and hole (-) branch of the dispersion.
Powerlaw scaling will extend out to
characteristic Kondo temperature $T_{K}$ of the system, so that the
total weight $x$ of f-electrons  contained within  the divergent peak is
$ 2x = \int_{-TK}^{T_{K}} N^{*} (E)\approx 4 \sqrt{T_{K}/T_{0}}$, giving
$T_{0}= 4 T_{K}/x^{2}$.

If the $f$-level is pinned to zero energy, then at
low temperatures a Fermi line of zero energy excitations forms along
the $c$-axis.  In a field, the Zeeman-splitting of the $f$-level induces
a singular polarization of nodal electron and hole bands, broadening
the Fermi line into a distinct tubular Fermi surface. 
When a field is introduced, a spin-polarized Fermi surface grows around the
line-zero in the hybridization, giving rise to a density of states of
order $N^{*} [\frac{g}{2}\mu_{B}B]\sim B^{-1/2}$, leading to a Pauli
susceptibility that diverges as $\chi \sim B^{-1/2}$.  We call this
field-induced Fermi surface transition a ``vortex transition''. Vortex
transitions are reminiscent of a Lifshitz
transition, but whereas Lifshitz transitions are point defects
in momentum space \cite{Dze,Voj}, the vortex transition is a line
defect.

We can model the singular thermodynamics of the system with the 
Free energy
\begin{eqnarray}\label{thefit}
F[B,T] &=& -T \sum_{\alpha =\pm 5/2}\int_{-\infty }^{\infty }
dE N (E)\ln  [1+e^{-\beta (E-g\mu_{B}B\alpha)}]
\cr&=& T^{3/2}\Phi\left(\frac{g\mu_{B}B}{T}\right)
\end{eqnarray}
where
\[
\Phi (y) = -\frac{1}{\sqrt{T_{0}}} \int_{0}^{\infty
}\frac{dx}{\sqrt{|x|}}\sum_{\alpha =\pm 5/2}\ln [1 + e^{-x- y\alpha  }]
\]
and $T_{0}^{-1/2}= (1/2)\sum_{\pm }T_{\pm}^{-1/2}$.
Fig.~\ref{figgy3} compares the experimental scaling curve
~\cite{Mat} with that
predicted by our simple model. 
However, while a qualitatively good fit to the observations is obtained
using a gyromagnetic ratio consistent with the single
ion properties of Yb in \ybal, 
the characteristic energy scale required to fit the
experimental results is $T_{0}\sim 6.5eV$, 
far greater than the 
characteristic Kondo temperature ($\sim 200K $) of this system \cite{Mat}. 
Using our relationship $T_{0}= 4T_{K}/x^{2}$, 
we can understand this scale by assuming that 
about $x\sim 0.1$ of the f-spectral weight is contained within the
vortex metal contribution to the density of states.

\figwidth=7cm
\figgy{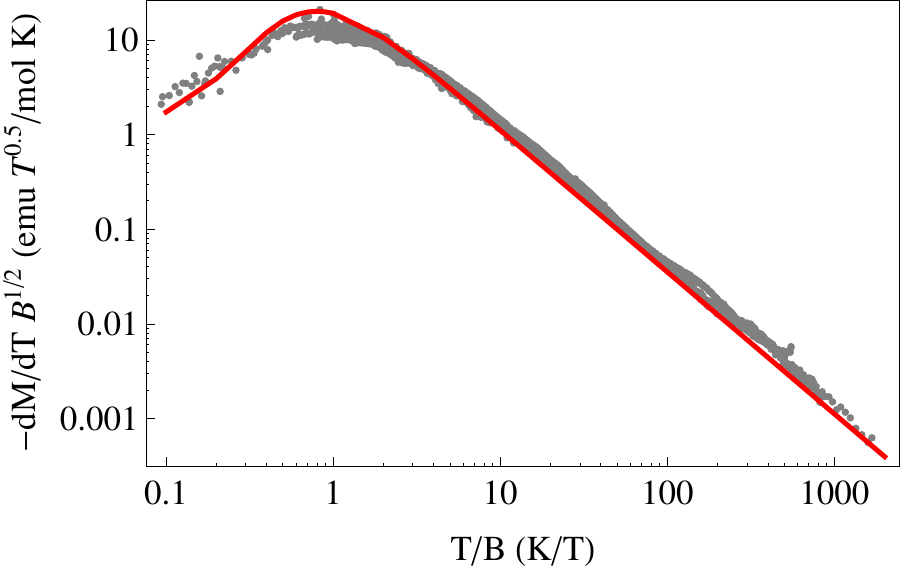}{Theoretical fit (red line) to the measured field-dependent
magnetization of \ybal from \cite{Mat} (gray dots) using formula \ref{thefit}
with $g m_{J}=2.85$ and $T_{0}= 6.65 eV$. }{figgy3}


We now turn to discuss some of the assumptions behind our model. 
One issue is whether the plane-wave description of the vortex metal 
survives inclusion of band-structure effects. In this
situation, angular momentum is only conserved
modulo n$\hbar $, where $n$ is the order of the symmetry group of the
Yb environment, requiring $n\ge 5$ to avoid any admixture
of $|m_{J}|=3/2, 1/2$  states into the perfect $\pm 5/2$ doublet.
In a model of 
\ybal, using tight-binding coupling within the B planes and
perfect heptagonal Yb rings, 
the nodal structure does indeed
survive, as shown in Fig. \ref{VLattice}. However more work is
required to understand whether the nodes persist in a more realistic
model of \ybal.
Another key assumption is that 
that the pinching of the hybridization gap by the node 
perfectly pins the $f$-level to the Fermi surface. 
Ultimately, this must arise from Coulomb screening, an 
effect that also needs inclusion in future work. 

Independent support for our phenomenological theory 
is provided by the locally isostructural polymorph $\alpha-$YbAlB$_4$,
which in contrast to the $\beta $ phase, has a
FL ground state \cite{Kug}.
Both systems display 
comparable characteristic ``Kondo'' scales
$T_K \approx 200K$ \cite{ybal-valence}. 
Recent experiments indicate in a
magnetic field,  \aybal develops a two
dimensional Fermi liquid at fields $B>3T$ \cite{OFarrell},
suggesting that \aybal is a phase in
which the $f$-level has become detached from the Fermi energy.
\figwidth=4.5cm
\figgy{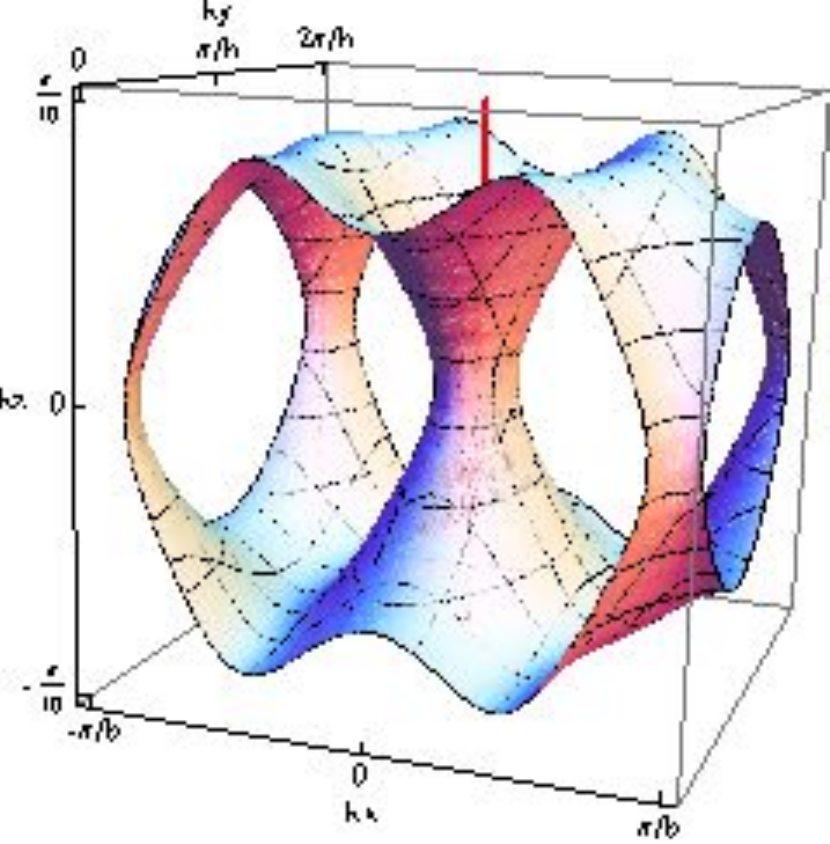}{Fermi surface from a 
tight-binding calculation. Note the cylindrical feature along the
z-axis linked to an almost 2D surface in the $k_{x}-k_{y}$ plane. For
clarity, the 1st BZ has been shifted by $\pi/h$ to move the node into
the center of the zone.}{VLattice}

Finally, we note that vortex structure in the
hybridization suggests a kind of topological line defect in momentum
space.  In Kondo insulators, the hybridization vanishes at the high
symmetry points forming point defects 
\cite{Dze},
corresponding to a homotopy ${\Pi}_{2} ({\cal H}) =
\mathbb{Z}_{2}$. Vortices in the hybridization suggest a further one
dimensional homotopy, $\Pi_1(H)=\mathbb{Z}$.
This is an interesting direction for future work.

Acknowledgments. The authors would like to thank Charles Kane, Alexey
A. Soluyanov, Rajif Roy, David Vanderbilt, Matthias Vojta and
especially Yosuke Matsumoto and Satoru Nakatsuji for discussions
related to their experimental work on \aybal and \ybal.  
The research was
supported by National Science Foundation Grants DMR-0907179 (AR and
PC) and 1066293 at the Aspen Center for Phyiscs (PC and AHN), DOE grant DE-AC02-98 CH 10886 (AMT) and the
David Langreth Graduate Development Award (AR).

\end{document}